\documentclass[aps,prl,twocolumn,superscriptaddress,showpacs,amsmath,amssymb]{revtex4-1}
\pdfoutput=1
\usepackage{graphicx}
\usepackage{dcolumn}
\usepackage{bm}

\newcommand{\equref}[1]{(\ref{#1})}
\newcommand{\mr}[1]{\mathrm{#1}}
\newcommand{\be}{\begin{equation}}
\newcommand{\ee}{\end{equation}}

\newcommand{\figta}{$\left(\mathrm{a}\right)\;$}
\newcommand{\figtb}{$\left(\mathrm{b}\right)\;$}
\newcommand{\figtc}{$\left(\mathrm{c}\right)\;$}
\newcommand{\figtd}{$\left(\mathrm{d}\right)\;$}

\newcommand{\figa}{$\left(\mathrm{a}\right)$}
\newcommand{\figb}{$\left(\mathrm{b}\right)$}

\newcommand{\figd}{$\left(\mathrm{d}\right)$}

\newcommand{\kohm}{\;\mr{k}\Omega}

\newcommand{\mhz}{\;\mr{MHz}}
\newcommand{\ghz}{\;\mr{GHz}}

\newcommand{\mk}{\;\mr{mK}}
\newcommand{\wb}{\;\mr{Wb}}
\newcommand{\nh}{\;\mr{nH}}

\newcommand{\mm}{\;\mr{mm}}

\newcommand{\cm}{\;\mr{cm}}

\newcommand{\mev}{\;\mr{meV}}
\newcommand{\ev}{\;\mr{eV}}

\newcommand{\nm}{\;\mr{nm}}
\newcommand{\kelvin}{\;\mr{K}}

\newcommand{\sloop}{S}
\newcommand{\bext}{B_{\mr{ext}}}
\newcommand{\phiext}{\Phi_{\mr{ext}}}
\newcommand{\fext}{f_{\mr{ext}}}
\newcommand{\lk}{L_\mr{k}}
\newcommand{\lgeom}{L_\mr{g}}
\newcommand{\phio}{\Phi_0}

\newcommand{\fq}{f_{\mr{q}}}
\newcommand{\fp}{f_{\mr{p}}}
\newcommand{\fs}{f_{\mr{s}}}
\newcommand{\fr}{f_{\mr{r}}}
\newcommand{\es}{E_\mr{S}}

\newcommand{\el}{E_\mr{L}}

\newcommand{\ip}{I_\mr{p}}

\newcommand{\dphi}{\delta\Phi}
\newcommand{\clen}{C_l}
\newcommand{\llen}{L_l}
\newcommand{\vph}{v}
\newcommand{\wave}{\bar{w}}
\newcommand{\wmin}{w_{\mr{min}}}
\newcommand{\ql}{Q_{\mr{L}}}

\renewcommand{\lgeom}{L_{\mr{g}}}
\newcommand{\lsq}{L_{\square}}
\newcommand{\rrq}{R_{\mr{Q}}}
\newcommand{\rsq}{R_{\square}}

\newcommand{\sigmaw}{\sigma_{w}}

\newcommand{\inox}{\mr{InO}_{\mr{x}}}

\begin{document}

\title{Coherent Flux Tunneling Through NbN Nanowires}

\author{J. T. Peltonen}
\email{joonas.peltonen@riken.jp}
\affiliation{RIKEN Center for Emergent Matter Science, Tsukuba, Ibaraki 305-8501, Japan}

\author{O. V. Astafiev}
\email{astf@zb.jp.nec.com}
\affiliation{RIKEN Center for Emergent Matter Science, Tsukuba, Ibaraki 305-8501, Japan}
\affiliation{NEC Smart Energy Research Laboratories, Tsukuba, Ibaraki 305-8501, Japan}
\affiliation{Department of Physics, Royal Holloway, University of London, Egham, Surrey TW20 0EX, United Kingdom}

\author{Yu. P. Korneeva}
\affiliation{Moscow State Pedagogical University, 01069, Moscow, Russia}

\author{B. M. Voronov}
\affiliation{Moscow State Pedagogical University, 01069, Moscow, Russia}

\author{A. A. Korneev}
\affiliation{Moscow State Pedagogical University, 01069, Moscow, Russia}
\affiliation{Moscow Institute of Physics and Technology, 141700, Dolgoprudny, Moscow Region, Russia}

\author{I. M. Charaev}
\affiliation{Moscow State Pedagogical University, 01069, Moscow, Russia}

\author{A. V. Semenov}
\affiliation{Moscow State Pedagogical University, 01069, Moscow, Russia}

\author{G. N. Golt'sman}
\affiliation{Moscow State Pedagogical University, 01069, Moscow, Russia}

\author{L. B. Ioffe}
\affiliation{Center for Materials Theory, Department of Physics and Astronomy, Rutgers University, 136 Frelinghuysen Road, Piscataway, New Jersey 08854, USA}

\author{T. M. Klapwijk}
\affiliation{Kavli Institute of Nanoscience, Delft University of Technology, Lorentzweg 1, 2628 CJ Delft, The Netherlands}

\author{J. S. Tsai}
\affiliation{RIKEN Center for Emergent Matter Science, Tsukuba, Ibaraki 305-8501, Japan}
\affiliation{NEC Smart Energy Research Laboratories, Tsukuba, Ibaraki 305-8501, Japan}

\date{\today}

\begin{abstract}
We demonstrate evidence of coherent magnetic flux tunneling through superconducting nanowires patterned in a thin highly disordered NbN film. The phenomenon is revealed as a superposition of flux states in a fully metallic superconducting loop with the nanowire acting as an effective tunnel barrier for the magnetic flux, and reproducibly observed in different wires. The flux superposition achieved in the fully metallic NbN rings proves the universality of the phenomenon previously reported for $\inox$. We perform microwave spectroscopy and study the tunneling amplitude as a function of the wire width, compare the experimental results with theories, and estimate the parameters for existing theoretical models.
\end{abstract}


\maketitle

{\it Introduction.}
Superconducting electrical circuits containing Josephson tunnel junctions have provided an ideal testing ground for investigating the quantum mechanics of macroscopic variables, starting with the observation of quantum coherence of the superconducting phase difference across a Josephson junction~\cite{martinis87} and leading to the development of superconducting qubits~\cite{clarke08}. Recently, it was realized that due to the fundamental charge--phase duality exhibited by Josephson devices, exactly dual physics can be observed in circuits containing narrow nanowires of highly disordered superconductors in which coherent quantum phase slips (CQPS) can have a significant probability amplitude~\cite{mooij06}. Thermally activated phase slips (PS) of the order parameter, corresponding to passage of a quantum of magnetic flux over the energy barrier represented by the wire, are a well-known origin of resistance below the critical temperature in superconducting wires~\cite{little67,tinkhambook,arutyunov08}. At the lowest temperatures, transport measurements indicate a transition to PS by incoherent quantum tunneling~\cite{giordano88,bezryadin00,altomare06,zgirski08}. Very recently CQPS was observed directly for the first time in strongly disordered $\inox$ nanowires embedded into superconducting loops~\cite{astafiev12}, demonstrating the concept of a PS flux qubit~\cite{mooij05}, dual to the single Cooper pair box~\cite{nakamura99}. However, several basic questions remain open, e.g., universality and reproducibility in different materials. Moreover, strongly disordered superconductors such as $\inox$ exhibit a number of properties different from conventional superconductors, in particular the role of dissipation~\cite{driessen12}, which make the study of QPS an interesting problem in itself.

\begin{figure}[!hb]
\includegraphics[width=\columnwidth]{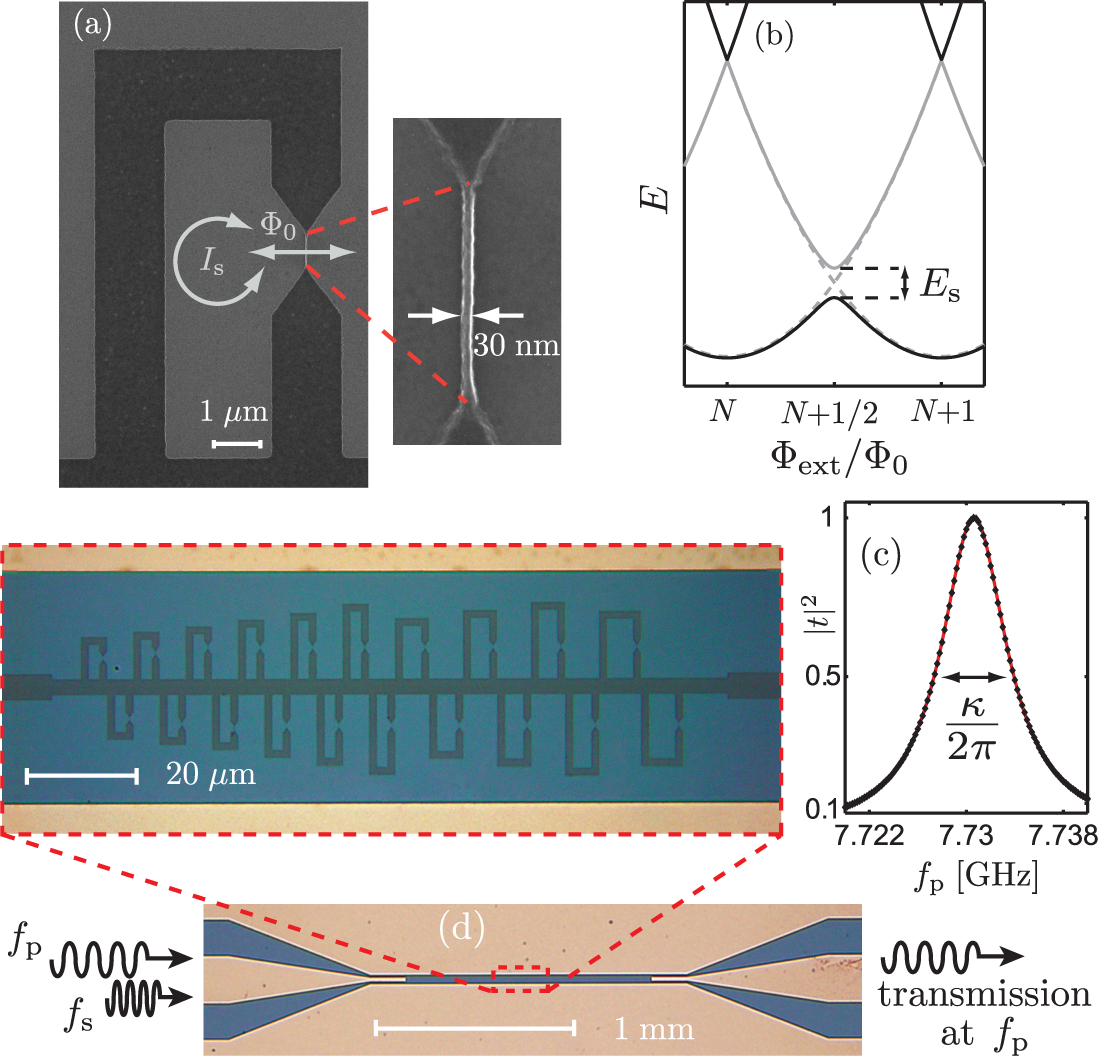}
\caption{(color online) \figta Scanning electron micrograph of a NbN PS flux qubit, illustrating the operation principle of the device. The nanowire is shown in a magnified view. \figtb Qubit energy levels in the limit $\es\ll\el$. The gray dashed lines show energies of the classical flux states. \figtc Measured resonator transmission (sample A) around the mode $f_3$ (black symbols), and a Lorentzian fit (solid red line). \figtd Optical microscope image of a typical sample, together with a schematic measurement diagram. The enlargement shows the center section with the 20 qubit loops.} \label{fig:scheme}
\end{figure}

In this Letter, we report the observation of coherent flux superpositions in fully metallic NbN loops, each containing a nanowire section as the tunnel barrier for magnetic flux (cf. Fig.~\ref{fig:scheme}). We observe the behavior in several loops on the same chip, characterize the dependence of the flux tunneling on the wire width, and compare the measurement results with the expected exponential dependence on the barrier width. Each of the two main findings of this work, (i) demonstration of coherent flux tunneling in a material different from $\inox$ and (ii) its wire-width dependence are of significant importance. They are crucial for developing more involved CQPS devices~\cite{hriscu11,kerman12,hongisto12,lehtinen12}, utilizing physics dual to conventional Josephson ones. Reproducing the flux superposition in the fully metallic superconducting rings shows that CQPS is a generic property of strongly disordered superconductors with large gap. Furthermore our results show an exponential dependence on the wire width that further proves the tunneling nature of the phase slip process which can be visualized as a virtual vortex crossing the wire. It is remarkable that such process that involves the rearrangement of many electrons remain nevertheless coherent.

{\it The device.}
The scanning electron micrograph of a typical loop in Fig.~\ref{fig:scheme}~\figta illustrates the working principle of a PS flux qubit~\cite{mooij05,mooij06,matveev02,arutyunov12}. A loop of NbN with nominal area $\sloop$ and high kinetic inductance $\lk$ is placed in a perpendicular magnetic field $\bext$. Due to flux quantization in superconducting loops~\cite{tinkhambook}, the total flux through the loop is an integer ($N$) multiple of the magnetic flux quantum $\phio=h/2e\approx 2\times 10^{-15}\wb$, and the energy of the loop is $E_{N}=\el(\fext-N)^2$, expressed in terms of the external flux $\fext=\phiext/\phio$ with $\phiext=\bext\sloop$ and the inductive energy $\el=\phio^2/2\lk$~\cite{lgnote}. The CQPS process in the nanowire, described by the amplitude $\es$, lifts the degeneracy of the fluxoid states $|N\rangle$ and $|N+1\rangle$ at $\phiext=(N+1/2)\phio$. The resulting energy band diagram is shown in Fig.~\ref{fig:scheme}~\figb, characterized by an avoided crossing of magnitude $\es$~\cite{mooij05}.

At $\phiext=(N+1/2)\phio$ the ground and first excited states correspond to symmetric and antisymmetric superpositions of $|N\rangle$ and $|N+1\rangle$, respectively. The energy splitting of this effective two level system is $h\fq=\sqrt{\varepsilon^2+\es^2}$. Here, $\varepsilon=2\ip\dphi$, with the persistent current $\ip=\phio/2\lk$ and $\dphi=\phiext-(N+1/2)\phio$, gives the difference $E_{N+1}-E_N$ away from the degeneracy. To probe $\fq$ and hence $\es$, we couple the loop to a coplanar NbN resonator via a section of shared kinetic inductance [bottom loop edge in Fig.~\ref{fig:scheme}\figa], enabling readout of multiple qubits located close to each other on a single chip~\cite{astafiev12}. We perform dispersive readout of the coupled qubit--resonator system by monitoring the amplitude and phase of transmitted microwaves~\cite{wallraff04} while varying $\phiext$.

{\it Experimental methods.}
Generally, the materials optimal for CQPS should be highly disordered and characterized by large normal state resistivity that translates into large impedance in superconducting state~\cite{mooij05}. At the same time this high degree of disorder should not suppress the superconducting gap or introduce subgap states as this would introduce dissipation and decoherence~\cite{astafiev12}. Transport data~\cite{gantmakher10,semenov01} in combination with STM measurements~\cite{sacepe08,sacepe10,sacepe11,noat12} indicate that materials favorable for CQPS include $\inox$, TiN, and NbN films.

Our samples were patterned from a NbN film of thickness $d\approx 2-3\nm$, deposited on a Si substrate by DC reactive magnetron sputtering~\cite{supplement}. The overview in Fig.~\ref{fig:scheme}~\figtd displays coplanar lines connecting to the external microwave circuit as well as the CPW resonator groundplanes. The resonator chip was enclosed in a sample box, and microwave characterization was performed in a dilution refrigerator at the base temperature of $40\mk$.

We focus on two out of several measured devices, fabricated simultaneously from the same film and cooled down at the same time, with identified qubits (two-level systems with transition controlled by microwave photons) belonging to 7 (10) out of the 20 loops for sample A (B), respectively. Referring to the enlarged view in Fig.~\ref{fig:scheme}~\figd, they are numbered from 1 to 20, starting from the smallest, i.e., the leftmost loop. The nominal wire width increases from $\gtrsim 20\nm$ in loop 1 to $\approx 75\nm$ in loop 20.

To characterize the qubits, we use a vector network analyzer and measure the complex microwave transmission coefficient $t$ through the resonator as a function of the frequency $\fp$ and the external field $\bext$. In addition, a second continuous microwave tone at $\fs$ can be used to excite the qubits through the resonator. The resonant modes are given by $f_n=n\vph/2L$, $n=1,2,3,\ldots$, where $L$ is the resonator length ($1.5\mm$ and $1.25\mm$ for sample A and B, respectively), $\vph=1/(\llen\clen)^{1/2}$ the effective speed of light, and $\llen$ ($\clen$) the inductance (capacitance) per unit length~\cite{supplement}. Figure~\ref{fig:scheme}~\figtc shows the squared amplitude of $t$ for sample A, at probing frequencies $\fp$ in a narrow range around $f_3=7.7306\ghz$, and normalized by the maximum transmission at $\fp=f_3$. A Lorentzian fit to the peak of $|t^2|$ gives the photon decay rate $\kappa=2\pi\times6.6\mhz$, corresponding to a loaded quality factor $\ql\approx 1.1\times 10^{3}$.

\begin{figure}[!htb]
\includegraphics[width=\columnwidth]{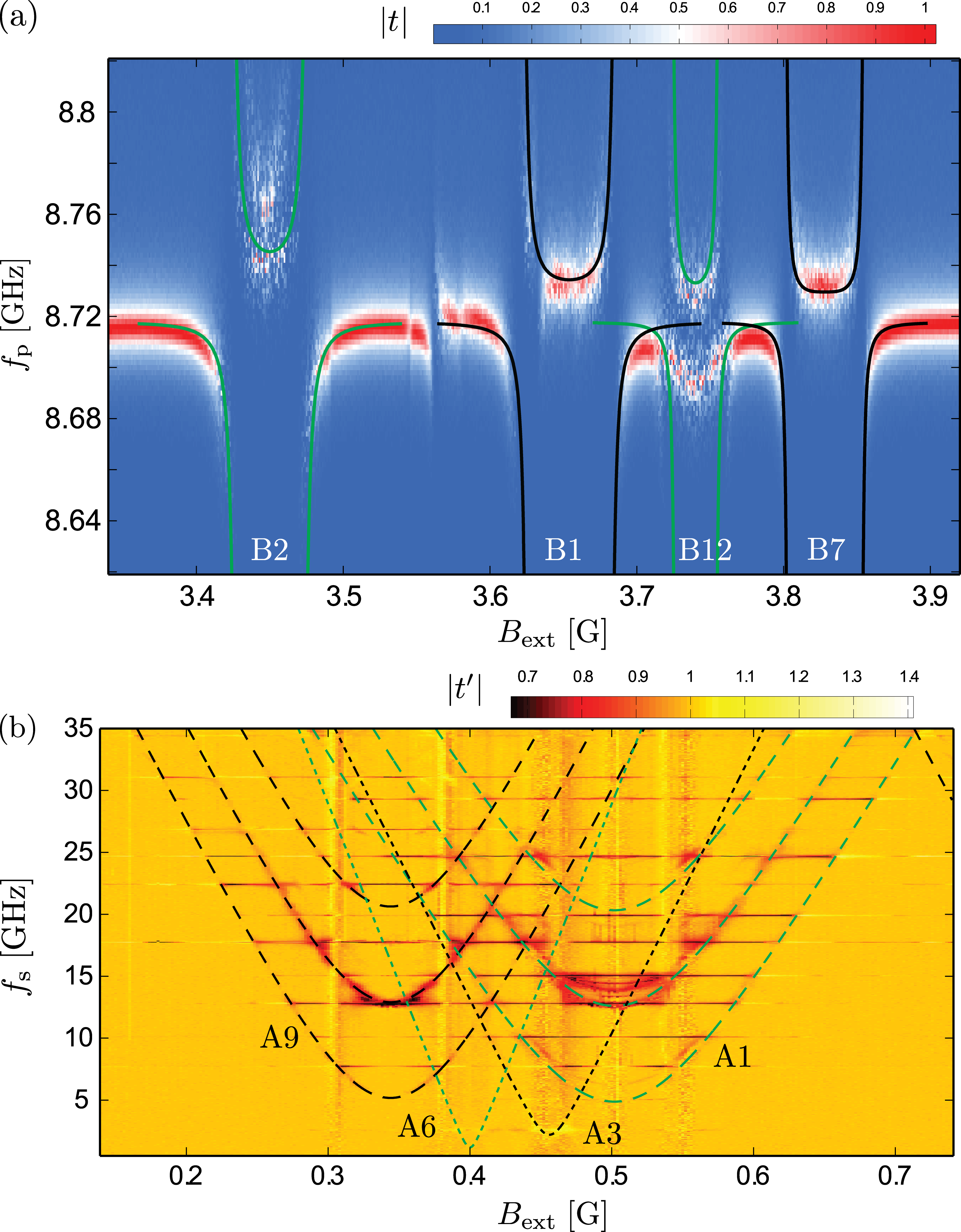}
\caption{(color online) \figta Amplitude of the normalized transmission coefficient t around the resonator mode $f_3$ (sample B). For four qubits, the lines show transition frequencies between the ground state and the two lowest dressed energy levels of the coupled qubit--resonator system. \figtb  Typical two-tone spectroscopy (sample A). The lines correspond to calculated qubit frequencies $\fq$ vs. $\bext$ for four qubits. The horizontal features originate from the resonator modes. Signatures of A6 and A3 are visible only close to the flux degeneracy points.} \label{fig:transspec}
\end{figure}

{\it Transmission measurements.}
Figure~\ref{fig:transspec}~\figta displays the result of the main qubit characterization measurement of sample B: $|t|$ in a range of $\fp$ around $f_3\equiv\fr$, and over a range of $\bext$. Avoided crossings typical for coherently coupled qubit--resonator systems are observed, with corresponding features present also in $\arg(t)$ (not shown). Measuring over a wider range of $\bext$ and extracting the periodicity in field of each feature in Fig.~\ref{fig:transspec}~\figta allows us to identify the loop from which they originate. Our calculations agree reasonably with the measured transmission~\cite{supplement}. For 4 qubits, the lines in Fig.~\ref{fig:transspec}~\figta show the two lowest transitions, calculated according to the Jaynes--Cummings model~\cite{wallraff04} by considering at a time only a single qubit coupled to the resonator.

To determine $\es$ and $\ip$ of the qubits (from the minimum value and slope of $\fq$ vs. $\bext$, respectively), we perform two-tone spectroscopy by continuously monitoring transmission at the fixed frequency $\fp=f_3$, while simultaneously sweeping the frequency $\fs$ of the additional spectroscopy tone over a wide range~\cite{astafiev10}. The result for sample A over a short range of $\bext$ is shown Fig.~\ref{fig:transspec}~\figb, including calculated $\fq(\bext)$ for selected qubits. $|t'|$ denotes the transmission amplitude normalized separately at each magnetic field by its value when $\fs$ is far detuned from any qubit or resonator transitions. The vertically offset curves with the same line type correspond to multiphoton processes with $\fs=\fq\pm\fp$. In some cases, telegraph noise typical for two-level fluctuations is observed. We attribute this to background charge fluctuators affecting $\es$.

\begin{table}[!htb]
\caption{\label{tab:sampletable}Qubit energies and wire widths.}
\begin{ruledtabular}
\begin{tabular}{lccccc}
\multicolumn{1}{c}{Loop} & $\wave\;[\mr{nm}]$ & $\wmin\;[\mr{nm}]$ & $\sigmaw\;[\mr{nm}]$ & $\es\;[\mr{GHz}]$ & $\es'\;[\mr{GHz}]$\footnotemark[1] \\
\hline
A1 & 27.4 & 21.6 & 2.3 & 12.6 &     \\
A2 & 26.8 & 20.2 & 2.6 & --   &     \\
A3 & 29.2 & 25.1 & 2.0 & 2.3  &     \\
A4 & 30.0 & 24.9 & 2.2 & 1.0  &     \\
A5 & 34.0 & 29.6 & 2.0 & --   &     \\
A6\footnotemark[2] & 31.5 & 27.2 & 1.9 & 0.9  &     \\
B1 & 28.0 & 22.2 & 2.4 & 7.0  & 7.0 \\
B2 & 29.6 & 23.2 & 3.0 & 7.3  & 5.5 \\
B3\footnotemark[3] & 29.0 & 24.1 & 1.7 & 1.4  & 0.9 \\
B4\footnotemark[3] & 29.1 & 24.8 & 2.2 & 0.8  & 1.0 \\
B5\footnotemark[3] & 30.7 & 26.8 & 1.9 & 1.6  & 2.5 \\
B6\footnotemark[2]\footnotemark[3] & 30.8 & 26.2 & 1.5 & --   & 1.3 \\
\end{tabular}
\end{ruledtabular}
\footnotetext[1]{Re-measurement of sample B after thermal cycling to $300\kelvin$}
\footnotetext[2]{Wire length $750\nm$ by design ($500\nm$ for wires 1--5); $\es$ normalized by 750/500}
\footnotetext[3]{$\es$ determined from $t$-measurement to approximately $\pm50\%$ accuracy (vs. $\lesssim 100\mhz$ with two-tone spectroscopy)}
\end{table}

\begin{figure}[!htb]
\includegraphics[width=\columnwidth]{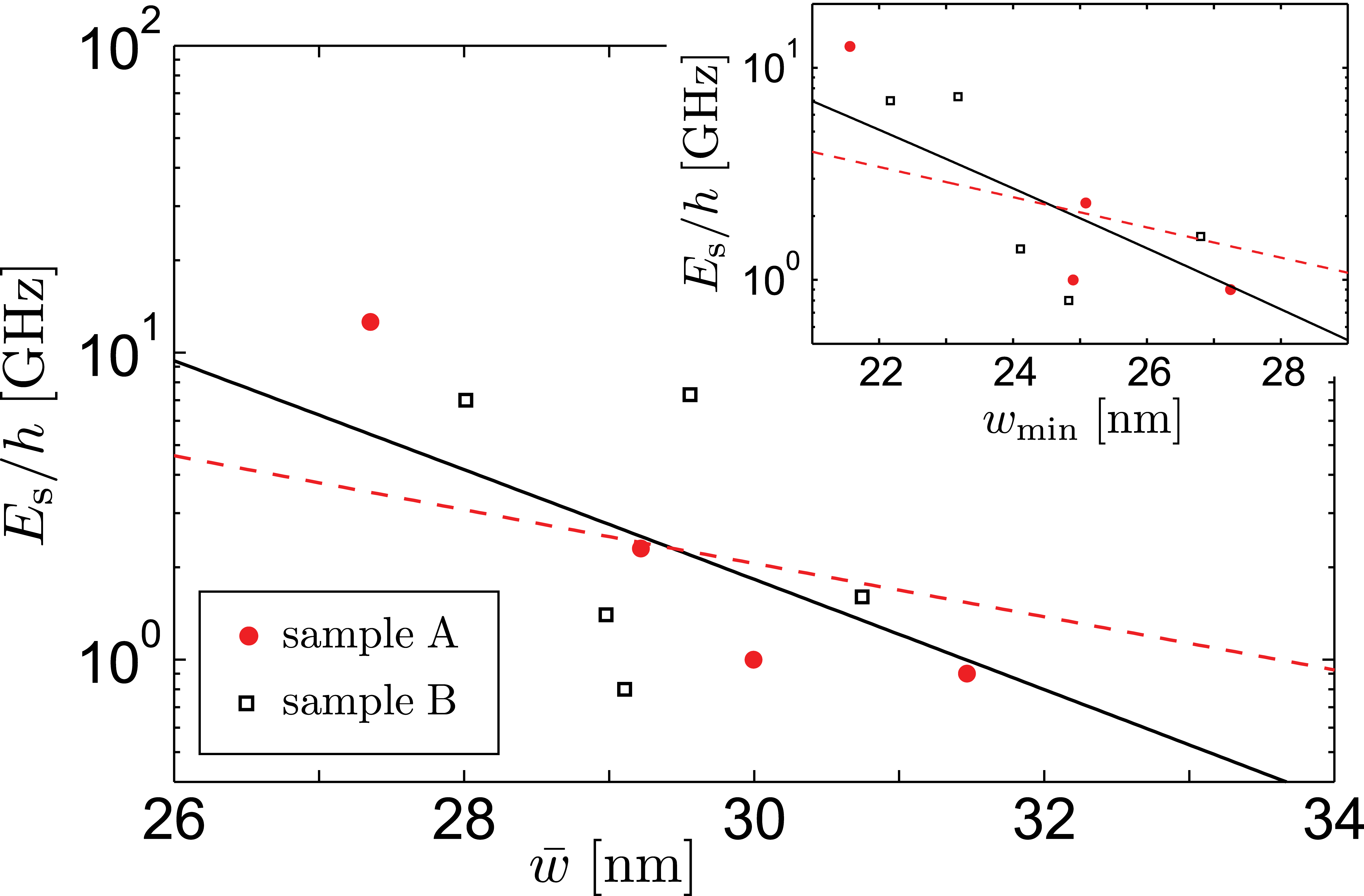}
\caption{(color online) Dependence of $\es$ on the average nanowire width $\wave$ extracted from SEM images by an automated procedure. Inset: $\es$ vs. $\wmin$. The symbols denote experimental data, and the lines are exponential fits (see text for details).} \label{fig:es}
\end{figure}

{\it Analysis of the phase slip amplitude.}
Table~\ref{tab:sampletable} and Fig.~\ref{fig:es} summarize the results. In Table~\ref{tab:sampletable} we collect the average wire widths $\wave$, the minimum widths $\wmin$, and the width standard deviations $\sigmaw$, together with the experimentally derived $\es$ and $\es'$, the latter obtained after thermal cycling of sample B to $300\kelvin$. Figure~\ref{fig:es} shows $\es$ versus $\wave$. For both samples, we focus on the qubits from loops 1--6 with wires of better quality (sample A: A1--A6 and B: B1--B6), featuring smallest relative roughness in width. During EBL, the nominally narrowest wires in these loops were written as single pixel lines, resulting in $\sigmaw\approx 2-3\nm$. In contrast, $\es$ of the other detected qubits (from loops 7--12, patterned in area mode with sub-optimal dose, yielding $\sigmaw\approx 6-8\nm$) do not follow any apparent dependence on $\wave$, indicating that these wires behave as multiply constricted rather than uniform barriers for the flux tunneling. We take the SEM resolution into account in the wire width derivation, while additional unknown systematic error can remain in the absolute values of $\wave$. Effective $\wave$ can also be reduced by a few nanometers due to oxidation at the edges. Nevertheless, it should not affect the overall dependence. Note that almost all wires 1--6 work as good tunnel barriers for the magnetic flux. However, signatures from loops A2 and A5 with minimal and maximal $\wave$ are not found. We suppose that this is due to too high (more than $15\ghz$) and too low (less than $0.5\ghz$) $\es/h$ to be detected by our methods, consistent with our expectations.

We now compare the data with the theoretical expectations. As any quantum tunneling, the phase slip process is expected to be exponential in the tunnel barrier width:
\be
E_{S}=E_{0}\exp(-\kappa\wave)\label{es1}
\ee
where $E_{0}/h$ is related to an attempt frequency and $\kappa^{-1}$ gives the width at which the wire becomes essentially a one dimensional channel characterized by large quantum fluctuations. Qualitatively, the trend in Fig.~\ref{fig:es} agrees with this exponential dependence. However, the $\es$--values exhibit large scatter. It can originate from small non-uniformities in material parameters or film thickness, or the remaining wire width roughness. In addition, because of the exponential dependence of the tunneling rate on the number of conduction channels $N_{\mr{ch}}$, mesoscopic fluctuations of the conductance~\cite{altshuler85} $\delta G\sim e^{2}/h$ are expected to result in large fluctuations $\delta\ln\es\sim\delta N_{\mr{ch}}\sim1$.

The BCS--based theory of QPS in moderately disordered superconductors~\cite{zaikin97,golubev01,arutyunov08} gives the parameters in Eq.~\equref{es1} for $\wave\lesssim\xi$: $E_0=\Delta(\rrq/\rsq)l\wave\xi^{-2}$ and $\kappa=a(\rrq/\rsq)\xi^{-1}$. Here, $\Delta$ is the superconducting energy gap, $\rrq=h/(4e^2)\approx6.4\kohm$ the quantum resistance, $\rsq$ the normal state sheet resistance of the film, $l=500\nm$ the wire length, $\xi$ the superconducting coherence length, and $a$ denotes a dimensionless parameter of order unity. We use $\Delta\approx1.6\pm0.1\mev$ inferred from direct measurements of the gap in NbN films similar to those used here, $\xi=4\nm$ known for thicker films~\cite{bell07}, and the approximate low temperature resistance $\rsq\approx 2\kohm$. A linear fit to $\ln(\es)$ yields the reasonable value $a\approx 0.6$ (solid black line in Fig.~\ref{fig:es}), whereas the corresponding kinetic inductance $\lsq=\hbar\rsq/\pi\Delta\approx 0.25\nh$ expected from BCS theory deviates from the measured $\lsq\approx 1.3\nh$. Poor applicability of the BCS theory, however, is not surprising for the strongly disordered material, and not strictly one-dimensional wires. Here also random charge distribution along the wire is not accounted, which results in $E_0\propto l$. Moreover, recent extension~\cite{semenov13} of the microscopic model~\cite{zaikin97,golubev01} indicates that interaction of individual phase-slip events can become relevant and affect the observable $\es$.

Now, we compute $\es$ according to the phenomenological model~\cite{ioffe10,feigelman10} of the strongly disordered superconductors, where the measured $\lsq$ enters directly as an input parameter. In this model $E_0=\rho\sqrt{l/\wave}$ and $\kappa=\eta\sqrt{\nu_{p}\rho}$, where $\rho=(\hbar/2e)^2\lsq^{-1}$ is the superfluid stiffness $(\rho/h\approx130\ghz)$, the numerical parameter $\eta\approx1$, and $\nu_{p}=1/(2e^2\rsq D)$ is the Cooper pair density of states~\cite{astafiev12,nupnote}. Based on the diffusion coefficient of the films $D\approx0.45\cm{}^{2}/\mr{s}$~\cite{semenov01} we fix $\nu_{p}\approx35\ev^{-1}\mr{nm}{}^{-2}$. A fit then yields the reasonable value $\eta\approx 1.4$ (dashed red line in Fig.~\ref{fig:es}). Next, in the inset of Fig.~\ref{fig:es} we show $\es$ as a function of $\wmin$. Assuming that $\es$ is dominated by the tunneling amplitude via a single constriction as suggested in Ref.~\onlinecite{vanevic12}, we approximate $l\approx\wmin$ and obtain $a\approx0.5$ (solid line) or $\eta\approx1.2$ (dashed). Note that estimates using $\eta=1$ give the correct order of the $\es$ without any fitting parameters.

Sample B was cooled down twice to study the effects of thermal cycling. As evident from Table~\ref{tab:sampletable}, $\es$ changes a little compared to the first measurement. This may be interpreted in terms of the Aharonov--Casher effect, i.e., interference of PS from different regions of the wire, and its dependence on the surrounding offset charges~\cite{manucharyan12,pop12}. As argued in Ref.~\onlinecite{vanevic12}, the PS nature of the wires is retained even if they contain weak constriction-type inhomogeneities: The requirement is that the constriction resistance is much smaller than the total wire resistance, a condition likely satisfied by our wires.

Besides the initial demonstration of CQPS in $\inox$ wires and the NbN wires discussed in this Letter, we have recently observed qubit behavior in nanowires from ALD--grown TiN as well as purposely-made short constrictions in NbN and TiN. Similar to $\inox$, the cause of strong decoherence in the nanowire qubits requires further study. For the fabrication of practical devices utilizing CQPS, the ideal would be a disordered material with highly reproducible fabrication process, together with minimized wire roughness. In conclusion, we find phase-slip flux qubit behavior with systematic wire-width dependence, in agreement with the theory of CQPS up to exponential accuracy.

\begin{acknowledgments}
The work was financially supported by the JSPS FIRST program and MEXT Kakenhi 'Quantum Cybernetics'. We acknowledge financial support from the Ministry of Education and Science of the Russian Federation (Agreement No. 14B.37.21.1214 and contract No. 14.B25.31.0007). L. B. I. acknowledges financial support from ARO W911NF-09-1-0395, ANR QuDec and John Templeton Foundation, and T. M. K. from EU MicroKelvin (No. 228464, Capacities Specific Programme), the Dutch Foundation for Research of Matter (FOM), and Ministry of Education and Science of the Russian Federation under contract No. 14.B25.31.0007. We thank E. F. C. Driessen and P. J. C. C. Coumou for helpful comments.
\end{acknowledgments}


\begin{thebibliography}{99}

\bibitem{martinis87} J. M. Martinis, M. H. Devoret, and J. Clarke, Phys. Rev. B {\bf 35}, 4682 (1987).

\bibitem{clarke08} J. Clarke and F. K. Wilhelm, Nature {\bf 453}, 1031 (2008).

\bibitem{mooij06} J. E. Mooij and Yu. V. Nazarov, Nature Phys. {\bf 2}, 169 (2006).

\bibitem{little67} W. A. Little, Phys. Rev. {\bf 156}, 396 (1967).

\bibitem{tinkhambook} M. Tinkham, {\it Introduction to Superconductivity}, 2nd ed. (McGraw-Hill, 1996).

\bibitem{arutyunov08} K. Yu. Arutyunov, D. S. Golubev, and A. D. Zaikin, Phys. Rep. {\bf 464}, 1 (2008).

\bibitem{giordano88} N. Giordano, Phys. Rev. Lett. {\bf 61}, 2137 (1988).

\bibitem{bezryadin00} A. Bezryadin, C. N. Lau, and M. Tinkham, Nature {\bf 404}, 971 (2000).

\bibitem{altomare06} F. Altomare, A. M. Chang, M. R. Melloch, Yu. Hong, and C. W. Tu, Phys. Rev. Lett {\bf 97}, 017001 (2006).

\bibitem{zgirski08} M. Zgirski, K.-P. Riikonen, V. Touboltsev, and K. Yu. Arutyunov, Phys. Rev. B {\bf 77}, 054508 (2008).

\bibitem{astafiev12} O. V. Astafiev, L. B. Ioffe, S. Kafanov, Yu. A. Pashkin, K. Yu. Arutyunov, D. Shahar, O. Cohen, and J. S. Tsai, Nature {\bf 484}, 355 (2012).

\bibitem{mooij05} J. E. Mooij and C. J. P. M. Harmans, N. J. Phys. {\bf 7}, 219 (2005).

\bibitem{nakamura99} Y. Nakamura, Yu. A. Pashkin, and J. S. Tsai, Nature {\bf 398}, 786 (1999).

\bibitem{driessen12} E. F. C. Driessen, P. C. J. J. Coumou, R. R. Tromp, P. J. de Visser, and T. M. Klapwijk, Phys. Rev. Lett. {\bf 109}, 107003 (2012).

\bibitem{hriscu11} A. M. Hriscu and Yu. V. Nazarov, Phys. Rev. B {\bf 83}, 174511 (2011).

\bibitem{kerman12} A. J. Kerman, arXiv:1201.1859 (2012).

\bibitem{hongisto12} T. T. Hongisto and A. B. Zorin, Phys. Rev. Lett. {\bf 108}, 097001 (2012).

\bibitem{lehtinen12} J. S. Lehtinen, K. Zakharov, and K. Yu. Arutyunov, Phys. Rev. Lett. {\bf 109}, 187001 (2012).

\bibitem{matveev02} K. A. Matveev, A. I. Larkin, and L. I. Glazman, Phys. Rev. Lett. {\bf 89}, 096802 (2002).

\bibitem{arutyunov12} K. Yu. Arutyunov, T. T. Hongisto, J. S. Lehtinen, L. I. Leino, and A. L. Vasiliev, Sci. Rep. {\bf 2}, 293 (2012).

\bibitem{lgnote} We neglect the contribution of the geometric inductance, estimated as $\lgeom\lesssim 0.003\lk$ for our films.

\bibitem{wallraff04} A. Wallraff, D. I. Schuster, A. Blais, L. Frunzio, R.-S. Huang, J. Majer, S. Kumar, S. M. Girvin, and R. J. Schoelkopf, Nature {\bf 431}, 162 (2004).

\bibitem{gantmakher10} V. F. Gantmakher and V. T. Dolgopolov, Phys.-Usp. {\bf 53}, 1 (2010).

\bibitem{semenov01} A. D. Semenov, G. N. Gol'tsman, and A. A. Korneev, Physica C {\bf 351}, 349 (2001).

\bibitem{sacepe08} B. Sac\'ep\'e, C. Chapelier, T. I. Baturina, V. M. Vinokur, M. R. Baklanov, and M. Sanquer, Phys. Rev. Lett. {\bf 101}, 157006 (2008).

\bibitem{sacepe10} B. Sac\'ep\'e, C. Chapelier,	T. I. Baturina,	V. M. Vinokur, M. R. Baklanov, and M. Sanquer, Nature Commun. {\bf 1}, 140 (2010).

\bibitem{sacepe11}  B. Sac\'ep\'e, T. Dubouchet, C. Chapelier, M. Sanquer, M. Ovadia, D. Shahar, M. V. Feigel'man, and L. B. Ioffe, Nature Phys. {\bf 7}, 239 (2011).

\bibitem{noat12} Y. Noat, T. Cren, C. Brun, F. Debontridder, V. Cherkez, K. Ilin, M. Siegel, A. Semenov, H.-W. H\"ubers, and D. Roditchev, Phys. Rev. B {\bf 88}, 014503 (2013).

\bibitem{supplement} See Supplemental Material for description of the sample fabrication process, resonator properties, and modeling of the microwave transmission.

\bibitem{astafiev10} O. V. Astafiev, A. A. Abdumalikov, Jr., A. M. Zagoskin, Yu. A. Pashkin, Y. Nakamura, and J. S. Tsai, Phys. Rev. Lett. {\bf 104}, 183603 (2010).

\bibitem{altshuler85} B. L. Altshuler, JETP Lett. {\bf 41}, 648 (1985).

\bibitem{zaikin97} A. D. Zaikin, D. S. Golubev, A. van Otterlo, and G. T. Zimanyi, Phys. Rev. Lett. {\bf 78}, 1552 (1997).

\bibitem{golubev01} D. S. Golubev and A. D. Zaikin, Phys. Rev. B {\bf 64}, 014504 (2001).

\bibitem{bell07} M. Bell, A. Sergeev, V. Mitin, J. Bird, A. Verevkin, and G. Gol'tsman, Phys. Rev. B {\bf 76}, 094521 (2007).

\bibitem{semenov13} A. G. Semenov and A. D. Zaikin, Phys. Rev. B {\bf 88}, 054505 (2013).

\bibitem{ioffe10} L. B. Ioffe and M. Mezard, Phys. Rev. Lett. {\bf 105}, 037001 (2010).

\bibitem{feigelman10} M. V. Feigel'man, L. B. Ioffe, and M. Mezard, Phys. Rev. B {\bf 82}, 184534 (2010).

\bibitem{nupnote} $\nu_{p}$ is related to the density of states of electrons per unit area $\nu_{\square}$ via $\nu_{p}=\nu_{\square}/2$.

\bibitem{vanevic12} M. Vanevic and Yu. V. Nazarov, Phys. Rev. Lett. {\bf 108}, 187002 (2012).

\bibitem{manucharyan12} V. E. Manucharyan, N. A. Masluk, A. Kamal, J. Koch, L. I. Glazman, and M. H. Devoret, Phys. Rev. B {\bf 85}, 024521 (2012).

\bibitem{pop12} I. M. Pop, B. Doucot, L. Ioffe, I. Protopopov, F. Lecocq, I. Matei, O. Buisson, and W. Guichard, Phys. Rev. B {\bf 85}, 094503 (2012).

\end{thebibliography}
\end{document}



\begin{center}{\large \bf Supplementary material}\end{center}
\vspace{0.5cm}

\renewcommand{\theequation}{S-\arabic{equation}}
\setcounter{equation}{0}
\renewcommand{\thefigure}{S-\arabic{figure}}
\setcounter{figure}{0}

{\bf COHERENT FLUX TUNNELING THROUGH NBN NANOWIRES} \vspace*{11pt} \\

J. T. Peltonen$^{1}$, O. V. Astafiev$^{1,2,3}$, Yu. P. Korneeva$^{4}$, B. M. Voronov$^{4}$, A. A. Korneev$^{4,5}$, I. M. Charaev$^{4}$, A. V. Semenov$^{4}$, G. N. Golt'sman$^{4}$, L. B. Ioffe$^{6}$, T. M. Klapwijk$^{7}$, and J. S. Tsai$^{1,2}$\\
$^{1}$ {\it RIKEN Center for Emergent Matter Science, Tsukuba, Ibaraki 305-8501, Japan} \\
$^{2}$ {\it NEC Smart Energy Research Laboratories, Tsukuba, Ibaraki 305-8501, Japan} \\
$^{3}$ {\it Department of Physics, Royal Holloway, University of London, Egham, Surrey TW20 0EX, United Kingdom} \\
$^{4}$ {\it Moscow State Pedagogical University, 01069, Moscow, Russia} \\
$^{5}$ {\it Moscow Institute of Physics and Technology, 141700, Dolgoprudny, Moscow Region, Russia} \\
$^{6}$ {\it Center for Materials Theory, Department of Physics and Astronomy, Rutgers University, 136 Frelinghuysen Road, Piscataway, New Jersey 08854, USA} \\
$^{7}$ {\it Kavli Institute of Nanoscience, Delft University of Technology, Lorentzweg 1, 2628 CJ Delft, The Netherlands} \\

\section{Sample fabrication process}
\label{sec:sample}

The samples were fabricated by first depositing a NbN film of thickness $d\approx 2-3\nm$ on a Si substrate by DC reactive magnetron sputtering. $d$ is determined by sputter-time using a calibration from thicker films. Thicker films (typically $4-6\nm$) deposited by the similar process are used for fabricating superconducting nanowire single-photon detectors, cf. Refs.~\cite{hadfield09,goltsman01,divochiy08}. Further details of the deposition process can be found in Refs.~\cite{goltsman03,korneev11}.

Starting from the uniform NbN film, the CPW resonator structure and the loops with nanowires were fabricated in a process involving two steps of electron beam lithography (EBL). In the first step, a standard bilayer of positive-tone resist was spin coated on the NbN film to form a mask for metal liftoff: EBL was used to pattern the coplanar resonator groundplanes as well as the transmission lines for connecting to the external microwave measurement circuit, and the metals were deposited in an electron beam evaporator ($5\nm$ Ti, $70\nm$ Au, and $10\nm$ Al on top).

During the second EBL step, the resonator center line and the embedded nanowire loops were patterned into a negative resist (MicroChem SU-8 2000.5 diluted 1:2 in cyclopentatone and spin coated at 7000 rpm) and finally transferred into the NbN film by reactive ion etching (RIE) in $\mr{CF}_4$ plasma. The resist remaining on top of the wires was not removed to protect the NbN wires from further oxidation, and the samples were stored under nitrogen atmosphere. Finally, the processed piece of wafer (typical size $2\cm\times2\cm$ with $3\times6$ resonator structures) was diced into individual $5\mm\times2.5\mm$ chips.

\section{Resonator properties}
\label{sec:resonators}

The top half of Fig.~1~\figtd shows the 20 NbN loops located around the center point of the resonator. The distance to the ground plane on either side of the center line is $20\mum$. The width of the center line is $5\mum$, with the exception of the $3\mum$ wide and $130\mum$ long middle section with the qubit loops. Both samples reported here contain a resonator with capacitive coupling, but NbN step impedance resonators similar to Ref.~11 in the main text performed comparably.  As a function of temperature, in both samples transmission through the resonator abruptly (dis-)appears at $T^*=4.9\pm 0.1\kelvin$.

As mentioned in the main text, the resonant modes are given by $f_n=n\vph/2L$, $n=1,2,3,\ldots$, where $L$ is the resonator length ($1.5\mm$ and $1.25\mm$ for sample A and B, respectively), $\vph=1/(\llen\clen)^{1/2}$ the effective speed of light, and $\llen$ ($\clen$) the inductance (capacitance) per unit length. The qubit loops couple stronger to the odd modes with maximum current close to the center of the resonator. Transmission measurements yield an average mode spacing of $2.43\ghz$ and $2.92\ghz$ for sample A and B, respectively. Using the estimate $\clen\approx 7\times10^{-11}\fm$, we find $\vph\approx 7.3\times 10^{6}\ms$ and $\llen\approx 2.7\times 10^{-4}\hm$, corresponding to the square inductance $\lsq\approx 1.3\nh$ and characteristic impedance $Z_1=(\llen/\clen)^{1/2}\approx 2\kohm\gg Z_0\approx 50\ohm$ of the line.

\section{Modeling the microwave transmission}
\label{sec:transmission}

To calculate the transmission amplitude $|t|$ through the resonator in the vicinity of a qubit--resonator crossing at $\wr\approx\wq$, we start from the total Hamiltonian of the coupled qubit--resonator system under microwave drive at $\wp$. We consider only a single qubit coupled to the resonator, and neglect the influence of the other qubits, assuming their frequencies to be far detuned from $\wr$. For a full quantitative comparison to the measured $t$ at $\bext$ values with overlapping features from several qubits, a more involved calculation with several qubits coupled to same resonator mode is required; see, e.g., Ref.~\cite{fink09}.

\begin{figure}[htb]
\begin{center}
\includegraphics[width=\columnwidth]{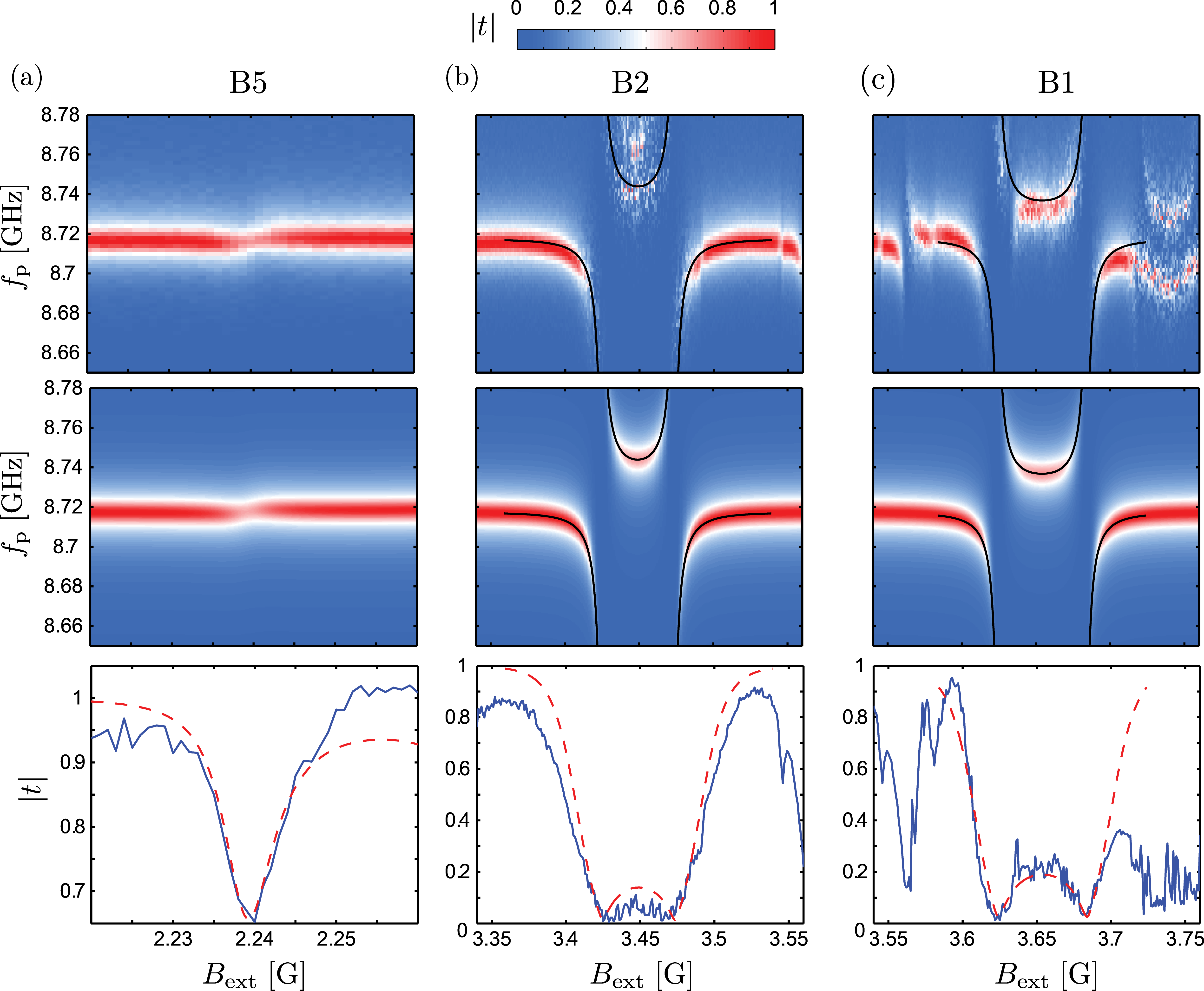}
\caption{\figb,\figtc Comparison of the measured microwave transmission to the prediction of Eq.~\equref{transmission} for qubits B2 and B1 of Fig.~2~\figa, respectively. The top panels show $|t|$ from the experiment, whereas the middle panels plot the corresponding modeled transmission amplitude. The bottom panels compare measured (solid) and calculated (dashed) horizontal line cuts along $\wp=\wr$, i.e., $\fp=f_3$. \figta Comparison of measured and calculated transmission of qubit B5 around $\wq\approx\wr$.}
\label{fig:tcompare}
\end{center}
\end{figure}

In a rotating wave approximation valid for small detunings $\delta_{q,r}=\omega_p-\omega_{q,r}$ we have~\cite{oelsner10}
\be
\mathcal{H}\approx-\frac{\hbar\delta_q}{2}\sigma_z-\hbar\delta_r a^{\dagger}a+\hbar g_{\varepsilon}(a^{\dagger}\sigma_{-}+a\sigma_{+})+\frac{\hbar\Omega}{2}(a^{\dagger}+a).\label{hamiltonian}
\ee
Here, $a^{\dagger}$ and $a$ are the photon creation and annihilation operators of the resonator, $\sigma_i$,~$\{i=x,y,z\}$ denote Pauli matrices in the qubit eigenbasis, and we define $\sigma_{\pm}=(\sigma_x\pm i\sigma_y)/2$. Further, $g_{\varepsilon}=g\es/(\hbar\wq)$ is the renormalized qubit--resonator coupling energy with $g=M\ip\ires$. Here, $M$ denotes the coupling inductance and $\ires$ the zero-point current fluctuation in the resonator. Finally, $\Omega$ is the amplitude of the microwave drive. For 4 qubits, the lines in Fig.~2~\figta show the two lowest transitions at frequencies $f_{\pm}=(\fq+\fr)/2\pm[(2g_{\varepsilon}/h)^2+(\fq-\fr)^2]^{1/2}/2$, determined by the eigenvalues of the non-driven Hamiltonian $(\Omega=0)$. They are indicated also by the black lines in Fig.~\ref{fig:tcompare}~\figtb and \figc.

To find the qubit dynamics we consider the master equation
\be
\dot{\rho}=\frac{1}{i\hbar}[\mathcal{H},\rho]+\mathcal{L}[\rho]\label{master}
\ee
for the density matrix $\rho$. Here, the Lindblad operator $\mathcal{L}=\mathcal{L}_r+\mathcal{L}_q$ takes into account two incoherent processes: Dissipation in the resonator (photon decay) at rate $\kappa$ is described by
\be
\mathcal{L}_r=\frac{\kappa}{2}(2a\rho a^{\dagger}-a^{\dagger} a\rho-\rho a^{\dagger}a),\label{lr}
\ee
assuming zero temperature for simplicity. Similarly, qubit decoherence is included via
\be
\mathcal{L}_q=\frac{\gamma_1}{2}(2\sigma_{-}\rho\sigma_{+}-\sigma_{+}\sigma_{-}\rho-\rho\sigma_{+}\sigma_{-})+\frac{\gamma_{\phi}}{2}(\sigma_z\rho\sigma_z-\rho),\label{lq}
\ee
describing relaxation at rate $\gamma_1$ and pure dephasing at $\gamma_{\phi}$.

In the limit of small driving amplitudes with the mean photon number in the resonator $\langle a^{\dagger}a\rangle\ll 1$, the photon space can be truncated to two states. As shown in Ref.~\cite{oelsner10}, under stationary conditions in this weak driving limit the transmission coefficient can be written in the compact form
\be
t=\frac{(i\delta_q'\kappa/2)}{(g_{\varepsilon}^2-\delta_r'\delta_q')}.\label{transmission}
\ee
Here, $\delta_q'=\delta_q+i\gamma$ and $\delta_r'=\delta_r+i\kappa/2$ denote the redefined detunings, and $\gamma=\gamma_{1}/2+\gamma_{\phi}$ is the total qubit dephasing rate.

We proceed to find $g_0=g/\ip=M\ires$ from the type of $|t|$ vs. $\bext$ measurements in Fig.~2~\figa, under the approximation it has the same value for all qubits. To achieve this, for each qubit clearly detected in two-tone spectroscopy, we first extract $\es$ and $\ip$ from the minimum value and slope of $\fq(\bext)$, respectively. Then, we use the total qubit decoherence rate $\gamma$ as an adjustable parameter and compare the measured $|t|$ with simulations and Eq.~\equref{transmission}. At the same time, the coupling $g_0$ is varied to obtain the best overall fit among all the qubits included in this procedure. The end result is exemplified by the comparisons for qubits B2 and B1 in Fig.~\ref{fig:tcompare}~\figtb and~\figc. The optimal value $g_0\approx8\mhz/\mr{nA}$ agrees with our expectations based on $M$ $(\lsq)$ and $\ires$.

Finally, we find $\es$ of the qubits not distinguishable from the low frequency noise floor in two-tone measurements (usually with $\es/h\lesssim 2\ghz$, cf. Table~1). In this case we treat $\es$ as an adjustable parameter along with $\gamma$ while using the above-determined value for $g_0$. An example comparison of the model and the measured transmission for qubit B5 is shown in Fig.~\ref{fig:tcompare}~\figa. For all the qubits, we obtain satisfactory agreement between the calculation and the measurements with total dephasing rates $\gamma/2\pi\approx 150-600\mhz$ depending on the qubit.

We observe $\ip$ to decrease as expected with increasing total number of squares of NbN film in a loop. Corresponding to $\ip$ extracted from spectroscopic and transmission measurements (probing local film properties), we find kinetic inductances in the range $\lsq\approx 1.3-1.6\nh$. These values are in reasonable agreement with $\lsq$ based on the resonator geometry and mode spacing (film properties on larger scale) and indicate the absence of large scale inhomogeneities in the film properties.